\documentclass[%
 reprint,
superscriptaddress,
nofootinbib,
 amsmath,amssymb,
 aps,
 prd,
floatfix,
]{revtex4-2}
\usepackage{graphicx}
\usepackage{dcolumn}
\usepackage{bm}
\usepackage{float}
\usepackage[usenames,dvipsnames]{color}
\usepackage{caption}
\usepackage{subcaption}
\usepackage{comment}
\usepackage{booktabs}
\usepackage{braket}
\usepackage{mathtools}
\usepackage[normalem]{ulem}

\begin{document}
\newcommand{\bea}{\begin{eqnarray}}
\newcommand{\eea}{\end{eqnarray}}
\newcommand{\beq}{\begin{equation}}
\newcommand{\eeq}{\end{equation}}
\newcommand{\sean}[1]{\textcolor{red}{\textbf{STM: #1}}}
\newcommand{\anuj}[1]{\textcolor{blue}{\textbf{#1}}}

\def\leq{\,\raise 0.4ex\hbox{$<$}\kern -0.8em\lower 0.62ex\hbox{$-$}\,}
\def\geq{\,\raise 0.4ex\hbox{$>$}\kern -0.7em\lower 0.62ex\hbox{$-$}\,}
\def\lsim{\,\raise 0.4ex\hbox{$<$}\kern -0.75em\lower 0.65ex\hbox{$\sim$}\,}
\def\gsim{\,\raise 0.4ex\hbox{$>$}\kern -0.75em\lower 0.65ex\hbox{$\sim$}\,}
\def\pm{\,\raise 0.4ex\hbox{$+$}\kern -0.75em\lower 0.65ex\hbox{$-$}\,}

\title{Direct Waves in Black-Hole Binary Mergers: Insights from the Backwards One Body Model}

\author{Anuj Kankani}
\email{anuj.kankani@mail.wvu.edu}
\affiliation{Department of Physics and Astronomy, West Virginia University, Morgantown, WV 26506, USA
}
\affiliation{Center for Gravitational Waves and Cosmology, West Virginia University, Chestnut Ridge Research Building, Morgantown, WV 26505,
USA
}
\author{Sean T. McWilliams}
\affiliation{Department of Physics and Astronomy, West Virginia University, Morgantown, WV 26506, USA
}
\affiliation{Center for Gravitational Waves and Cosmology, West Virginia University, Chestnut Ridge Research Building, Morgantown, WV 26505,
USA
}

\date{\today}

\begin{abstract}
The merger-ringdown radiation from a black hole binary merger is accurately modeled by a sum of linear quasinormal modes (QNMs). Recently, the ``direct wave" component of the radiation, associated with prompt emission from a plunging perturber, has been identified. Motivated by the behavior of null geodesics perturbed from the remnant light ring, the Backwards One Body (BOB) approach has been shown to model the full merger-ringdown radiation to high accuracy, while using only a minimal number of parameters. In this work, using the P\"oschl--Teller potential, we first show how the BOB amplitude evolution can be recovered from the QNM pole contributions. We then apply rational filters to BOB and numerical relativity waveforms. We show that the BOB rational filter residual agrees well with the NR rational filter residual, possibly helping explaining its accuracy near the waveform peak. Finally, we show that the direct wave frequency follows a different spin dependence than the horizon frequency, even for high spin remnants, and instead tracks the News frequency at the time of the peak News amplitude.
\end{abstract}

\maketitle

\section{Introduction}
The merger and ringdown radiation emitted from a black hole binary merger provides a direct probe for testing General Relativity (GR) in the strong field regime \cite{grtest1,grtest2,grtest3,grtest4,grtest5}. The ringdown phase is described by black hole perturbation theory \cite{regge1957stability, vishveshwara1970scattering, teukolsky1972rotating, kokkotas1999quasi, berti2009quasinormal,ferrari} and can be accurately modeled as a sum of linear quasinormal modes \cite{giesler2019black,cheung2024extracting,pacilio2024flexible,baibhav2023agnostic,magana2025high}. Additionally, there has been evidence for the presence of non-linear QNMs in numerical relativity (NR) waveforms \cite{mitman2023nonlinearities, cheung2023nonlinear, ma2024excitation, zhu2024nonlinear, redondo2024spin, bucciotti2024quadratic}, as well as late time tails \cite{ma2025merging, de2025late, islam2025phenomenology}. Recently, a distinct non-QNM contribution to the merger radiation, the ``direct wave'' \cite{oshita2025probing}, was identified through the application of rational filters \cite{ma2022quasinormal, ma2023black, ma2023using, lu2025statistical}, which allow for the removal of QNM content without requiring any fitting.

Another approach to modeling the merger-ringdown radiation is the Backwards One Body (BOB) model \cite{mcwilliams2019analytical, kankani2025bob, kankani2025modeling, mahesh2025combining}. Based on the behavior of a packet of null geodesics perturbed from the light ring of the remnant black hole \cite{ferrari,mcwilliams2019analytical}, BOB produces a hyperbolic-secant amplitude evolution and has been shown to model the NR merger-ringdown News (i.~e.~the first time derivative of the strain) with high accuracy for quasi-circular and non- or mildly-precessing cases across the SXS catalog, while requiring no additional parameters beyond what would already be needed to calibrate a single QNM \cite{mcwilliams2019analytical,kankani2025bob}. Although BOB is distinct from a sum of QNMs, it shares several important features with QNM-based descriptions. In particular, at late times its amplitude evolution approaches that of the fundamental QNM, leading to analytic relations between the peak News amplitude and the fundamental mode excitation \cite{ma2021universal,kankani2025bob}. Importantly, BOB performs significantly better than a QNM sum prior to the peak while using far fewer free parameters \cite{kankani2025bob}.

Despite its empirical success, the physical origin of BOB's remarkable accuracy near the waveform peak remains unclear. In this Letter we clarify the origin of this behavior. First, using the P\"oschl--Teller potential \cite{poschl1933bemerkungen}, we show that the QNM pole contributions produce the hyperbolic secant amplitude profile predicted by BOB. Then, applying rational filters to both BOB and NR waveforms, we demonstrate that BOB accurately models the rational filter residual. Taking advantage of BOB's analytical form, we explore properties of the direct wave frequency across the parameter space and show that even for high spin systems, it follows a different spin dependence than the horizon frequency of the remnant black hole. Rather, we find a strong correlation with the News frequency at the time of the peak News amplitude.

\section{A QNM approach to BOB}
While BOB shares many similarities to a sum of linear quasinormal modes, it only directly matches a QNM description at late times, when only the fundamental mode remains \cite{kankani2025bob, ma2021universal}. BOB matches the accuracy of 4--5 QNMs at the peak of the waveform, while requiring significantly fewer free parameters, and retains better behavior before the peak \cite{kankani2025bob}. The original BOB derivation was based on the divergence of null rays from the remnant black hole \cite{ferrari,mcwilliams2019analytical}. To gain further intuition into the connection between BOB and QNMs, we follow a more standard QNM derivation procedure based on the Sasaki--Nakamura (SN) formalism \cite{sasaki1982gravitational,berti2006quasinormal}. We only provide a brief overview of the standard QNM portion of the derivation, referring the reader to \cite{berti2006quasinormal} for more details. Perturbations induced by a spin-$s$ field are described by a function $X^{(s)}(t,r)$ whose Laplace transform satisfies
\begin{equation}
    \frac{d^2 \hat{X}^{(s)}(\omega,r)}{dr^2_*} + V_{\text{eff}}(r)\,\hat{X}^{(s)}(\omega,r) = I(\omega,r),
    \label{eq:sn_inhom_eq}
\end{equation}
where $r_*$ is the tortoise coordinate, $V_{\text{eff}}(r)$ is an effective potential, and $I(\omega,r)$ represents the initial data. Assuming the observer is located far from the black hole, we obtain the standard description of the ringdown as a sum of quasinormal modes \cite{berti2009quasinormal}.
\begin{equation}
X^{(s)}(t,r) =-\operatorname{Re}\left[\sum_{n}C_n e^{-i\omega_n(t-r_*)}\right],
\label{eq:BHPT_QNM}
\end{equation}
where $C_n$ are quasinormal excitation coefficients, represented as the product of the quasinormal excitation factor $B_n$ and an integral over the source $I_n$, $C_n = B_nI_n$. Additional details are provided in the appendix. Because the exact form of the initial data sourcing QNMs for astrophysically realistic mergers is unknown, in practice the QNM amplitudes are obtained by fitting to a known waveform. The BOB approach to merger-ringdown modeling suggests that the effect of the initial data is captured by the overall peak amplitude, $A_p$. Furthermore, several studies analyzing NR waveforms have found evidence that the source term for prograde QNMs is largely independent of the overtone index $n$ \cite{mitman2025probing,giesler2019black,cheung2024extracting,oshita2021ease}. This assumption greatly simplifies the analytical study of the merger-ringdown radiation and allows us to gain insight into why BOB successfully models the merger-ringdown radiation.
Under the assumption that prograde QNMs have, to a good approximation, a common source term, we can rewrite Eq.~(\ref{eq:BHPT_QNM}) as 
\begin{equation}
    X^{(s)} (t,r) \propto \operatorname{Re}\bigg[\sum_n B_n e^{-i\omega_n(t-r_*)}\bigg].
\end{equation}
We employ the P\"oschl--Teller (PT) potential \cite{poschl1933bemerkungen}, a common substitution for the true Regge--Wheeler potential \cite{ferrari,kuntz2025green,berti2006quasinormal,berti2009quasinormal,konoplya2011quasinormal} due to its simpler analytic structure. 
\begin{equation}
    V_{\text{PT}} = \frac{V_0}{\cosh^2\alpha(x-x_0)},
\end{equation}
where $x_0$ is the location of the maximum, $V_0$ is the height of the potential and $\alpha$ is related to the second derivative of $V_{\text{PT}}$. Because we are primarily concerned with the behavior of the potential around the peak, where the PT potential can be tuned to mimic the true effective potential, it is an accurate approximation for our calculation. Based on the reflection and transmission coefficients of the PT potential, calculated in \cite{ferrari}, the quasinormal excitation factors $B_n$ were calculated explicitly in \cite{berti2006quasinormal}\footnote{There is a sign error in appendix A3 of \cite{berti2006quasinormal} due to the opposite sign convention for $\omega$ in \cite{ferrari}. The $B_n^{\text{PT}}$ expression remains unchanged, but the correct pole condition is $-\beta -i\omega/\alpha = -n$.} as
\begin{equation}
    B_n^{\text{PT}} = \frac{i\alpha (-1)^{n+1} \Gamma(n-\beta) \Gamma(1+2\beta - n)}{2\omega_n n! \Gamma(1+\beta)\Gamma(-\beta) \Gamma(\beta - n)},
\end{equation}
\\\newline where $\beta = -1/2 + \sqrt{1/4 - V_0/\alpha^2}\equiv -1/2 + i\delta$.
The QNM can be identified from the poles in the reflection and transmission coefficients as $\omega_n = i\alpha \sqrt{1/4 - V_0/\alpha^2}  - i\alpha(n+1/2)$ \cite{berti2006quasinormal,ferrari}. Based on comparisons to NR waveforms, \cite{kankani2025bob} argued that the BOB amplitude evolution best represents the gravitational wave News, $\mathcal{N}$, while $X^{(s)}$ is more directly related to the gravitational wave strain, $h$. Because the News is the first time derivative of the strain, differentiating the above result and substituting $B_n^{\text{PT}}$ results in
\begin{align}
    &\frac{dX^{(s)} (t,r)}{dt}  \propto \operatorname{Re}\bigg[\frac{\alpha}{2\Gamma(1+\beta)\Gamma(-\beta)}\nonumber\\ 
    &\bigg[\sum_n \frac{(-1)^n \Gamma(n-\beta)\Gamma(1+2\beta - n)}{n! \Gamma(\beta-n)} e^{-i\omega_n(t-r_*)}\bigg]\bigg].
\end{align}
Employing the identities $\Gamma(z-n) = (-1)^n\Gamma(z)/(1-z)_n$ and $(x)_n = \Gamma(x+n)/ \Gamma(x)$ where $(x)_n$ is the Pochhammer symbol, allows us to rewrite our series as a hypergeometric function \cite{NIST:DLMF15.2}:

\begin{align}
    &F(-\beta,-1-\beta,-2\beta;e^{-\alpha(t-r_*)}) \nonumber\\=& \sum_{n=0}^\infty \frac{(-\beta)_n (-1-\beta)_n}{(-2\beta)_n}\frac{e^{-n\alpha(t-r_*)}}{n!}.
\end{align}

Using the identity $F(a,b;c;z) = (1-z)^{c-a-b}F(c-a,c-b;c;z)$, we obtain
\begin{widetext}
\begin{align}
    \frac{dX^{(s)} (t,r)}{dt} &\propto \operatorname{Re} \bigg[\frac{\alpha \Gamma(1+2\beta)}{2\Gamma(\beta) \Gamma(1+\beta)}e^{+\alpha \beta (t-r_*)} (1+e^{-\alpha(t-r_*)})^{-1} F\bigg(-\beta,-1-\beta;-2\beta;-e^{-\alpha(t-r_*)}\bigg)\bigg].\nonumber\\
    &\propto \operatorname{Re} \bigg[\text{sech}\bigg(\frac{t-r_*}{\tau}\bigg)e^{i\alpha \delta (t-r_*)}
    F\bigg(\frac{1}{2}-i\delta,\frac{-1}{2} - i\delta; 1-2i\delta;-e^{-\alpha(t-r_*)}\bigg) \bigg]
    \label{eq:dX_dt_propto_sech}
\end{align}
\end{widetext}
where $\tau = 2/\alpha$ is the inverse of the imaginary part of the fundamental QNM. Converting the master variable $dX^{(s)}/dt$ to the News, we find for the PT potential
\begin{equation}
|\mathcal{N}_+(t) - i\mathcal{N_\times}(t)| \propto \text{sech}\bigg(\frac{t-t_p}{\tau}\bigg)|F(t)|,
\end{equation} 
under the assumption that for non-spinning systems in the eikonal limit $c_0 \propto l^2$ \cite{sasaki1982gravitational,berti2006quasinormal,Sasaki_strain}. Here we have suppressed the $(l,m)$ dependence, $F(t)$ is the specific hypergeometric function appearing in Eq.~(\ref{eq:dX_dt_propto_sech}), and we have used $t-r_* \equiv t-t_p$. We provide greater detail on the conversion from $X^{(s)}$ to $\mathcal{N}$ in the appendix. 

\begin{figure}[h]
    \centering
    \includegraphics[]{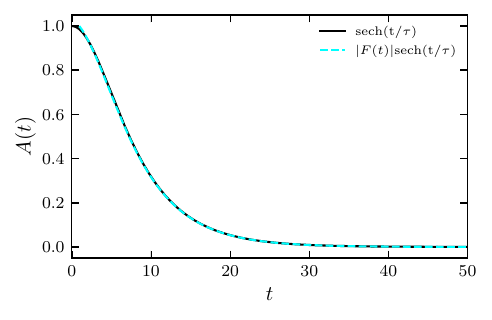}
    \caption{Comparison of sech$(t/\tau)$ (solid black) and $|F(t)|\text{sech}(t/\tau)$ where $F(t)$ is the hypergeometric function appearing in Eq.~(\ref{eq:dX_dt_propto_sech}).}
    \label{fig:hypergeometric} 
\end{figure}

Noting that the hypergeometric function appearing in Eq.~(\ref{eq:dX_dt_propto_sech}) is a sum of exponentially damped terms, we treat $|F(a,b;c;-e^{-\alpha(t-r_*)})|$ as a constant. In Fig.~(\ref{fig:hypergeometric}) we show the pure hyperbolic secant profile predicted by BOB, $\text{sech}(t/\tau)$, and the hypergeometric corrected profile, $|F(-\beta,-1-\beta,-2\beta;e^{-\alpha(t-r_*)})|\text{sech}(t/\tau)$, with the PT potential tuned to mimic the Zerilli potential, $x_0 = 0.95, V_0 = 0.605$ and $\alpha = 0.362$ \cite{kuntz2025green}. Both profiles are normalized to a maximum value of 1, due to the freedom allowed by the overall normalization parameter $A_p$ present in BOB, and the hypergeometric profile is timeshifted by $x_0$, corresponding to the shift in the peak of the PT potential. The two profiles show excellent agreement, with a very small disagreement near the peak. While this shows that treating $|F|$ as a constant does not significantly impact our amplitude profile, this derivation suggests that a hypergeometric correction is physically motivated and will likely lead to more accurate corrections to BOB in the future.

This approximation results in an amplitude profile exactly matching that from the original derivation based on the perturbations of null geodesics from the light ring.
\begin{equation}
|\mathcal{N}(t)| \propto \text{sech}\bigg(\frac{t-t_p}{\tau}\bigg),
\label{eq:news_propto_sech}
\end{equation}
Importantly, reproducing the hyperbolic secant profile involves $dX^{(s)}/dt$, rather than $X^{(s)}$, complementing the empirical evidence provided in \cite{kankani2025bob} that BOB fundamentally models the News, rather than $\Psi_4$ as was originally suggested in \cite{mcwilliams2019analytical} \footnote{Although, as has been shown in \cite{mcwilliams2019analytical} and \cite{kankani2025bob}, the BOB amplitude and frequency profile can still model $\Psi_4$ accurately.}. While we have limited ourselves to the PT potential, the success of BOB across the SXS catalog \cite{kankani2025bob}, where the Kerr $(s=-2,l=2,m=2,n=0)$ QNM is used, suggests that physical insight gained in the simpler PT potential translates well to the full Kerr potential.

\section{Studying BOB with Rational Filters} The above derivation shows that BOB's approach of modeling the merger-ringdown radiation as a perturbation of null geodesics at the remnant light ring correctly incorporates the linear QNM contribution, in the eikonal limit for overtones, to the amplitude. Importantly, BOB does not directly model the dynamics of the perturber, but instead models the dynamics of the radiation. The radiation that is generated as the perturber crosses the light ring should follow dynamics similar to the prompt radiation emitted immediately before the crossing. Near the peak of the potential, both the prompt and QNM radiation are primarily governed by the effective potential near its peak. We therefore expect BOB to capture more than the linear QNM content of the merger-ringdown radiation. This also explains why BOB outperforms a sum of QNMs, on a free parameter basis, near and before the peak of the waveform \cite{kankani2025bob}. We test this line of reasoning through the usage of rational filters.

Rational filters \cite{ma2022quasinormal,ma2023using,lu2025statistical}, which work in the frequency domain, help remove QNM content from merger-ringdown waveforms without requiring any fitting. As shown in the derivation above, BOB's amplitude evolution correctly incorporates information from an infinite number of overtones in the eikonal limit. Rational filters were employed in \cite{oshita2025probing} to identify the presence of a distinct non-QNM component of the merger radiation, the ``direct wave'', which was associated with the direct emission from the plunging perturber \cite{oshita2025probing}.

We focus our analysis on three quasicircular and non-precessing NR simulations. The first, SXS:BBH:0305 \cite{SXS:BBH:0305,SXSCatalogData_3.0.0,Boyle2025SXS,sxs_cat3}, has parameters similar to GW150914 \cite{gw150914}, and was analyzed with rational filters in \cite{ma2023using, oshita2025probing}. The second, SXS:BBH:4123 \cite{Boyle2025SXS,SXSCatalogData_3.0.0,sxs_cat3,SXS:BBH:4123}, consists of a $q=4$ mass ratio system with a highly spinning remnant, $\chi_f = 0.915$. The last simulation, SXS:BBH:2477 \cite{Boyle2025SXS,SXSCatalogData_3.0.0,sxs_cat3,Yoo_2022,SXS:BBH:2477}, consists of a $q=15$ mass ratio system with a low spin remnant, $\chi_f = 0.189$. Because BOB most accurately models the gravitational wave News \cite{kankani2025bob}, we apply rational filters to the News rather than the strain. The News contains the same QNM frequencies as the strain; therefore, the same rational filters apply. For the NR simulations, following \cite{oshita2025probing}, we remove 7 prograde, 2 retrograde and the (3,2) mode. We obtain the NR News through a cubic spline differentiation of the NR strain. For BOB, we employ two approaches. The first is to remove the exact same QNMs from BOB that we do from the NR waveform. In the second approach, we account for how BOB incorporates overtone information into its amplitude evolution. The BOB amplitude evolution can be rewritten as 
\begin{align}
    A_p\text{sech}\left(\frac{t-t_p}{\tau}\right) &= 2 A_p \sum_{n=0}^{\infty} (-1)^n e^{-(2n+1)\frac{t-t_p}{\tau}}\nonumber\\ 
    &= 2A_p\bigg(e^{-\frac{t-t_p}{\tau}} - e^{-3\frac{t-t_p}{\tau}} + e^{-5\frac{t-t_p}{\tau}} + ...\bigg),
    \label{eq:amplitude_expansion}
\end{align}

where $A_p$ is the peak amplitude of the waveform, providing an overall normalization\footnote{An extended discussion of the frequency and phase evolution can be found in \cite{kankani2025bob}.}. 
From this, we can see that BOB incorporates the damping times of overtones through the eikonal limit where $\tau_n = \tau_{n=0}/(2n+1)$. We note that BOB cannot be matched term by term to a sum of QNMs until late times. Rather it produces the net result of an infinite number of QNMs, in the eikonal limit. In our second approach to filtering BOB, we only remove seven prograde QNMs. The imaginary part of the overtones is incorporated through the eikonal limit. The fundamental mode, and the real part of the QNMs, use the values determined from the \texttt{qnm} package \cite{stein2019qnm}.

Lastly, we note that by rewriting $(-1)^{n}$ as $e^{{i\pi}n}$ in Eq.~(\ref{eq:amplitude_expansion}), we can describe the News, $N(t) = A(t) e^{i\Phi(t)}$, for BOB as
\begin{equation}
    \mathcal{N}(t) = 2A_p\bigg(e^{-\frac{t-t_p}{\tau} + i\Phi(t)} + e^{-3\frac{t-t_p}{\tau} + i(\pi + \Phi(t))} + ...\bigg)
    \label{eq:News_as_sum}
\end{equation}
where $\Phi(t)$ is the BOB phase evolution, described in \cite{kankani2025bob}. Since each term in Eq.~(\ref{eq:News_as_sum}) has the same $\Phi(t)$, each term in the series has a phase that is offset by $\pi$ relative to the previous term, so that the terms in the series have alternating signs. Recent studies investigating the QNM content in NR data, including through the use of rational filters and analytically in the Schwarzschild spacetime, have found that successive overtones are excited with phases that have terms with alternating signs \cite{giesler2019black,ma2022quasinormal,coleman2025high,deamicis2026}. While BOB's evolution can not be correlated exactly term by term to a multi-QNM model, BOB's correct accounting of QNM information, in the eikonal limit for overtones, as seen in the previous section, results in it naturally incorporating the excitation features seen in NR waveforms, such as the alternating-sign phase structure for overtones.

\begin{figure}[h]
    \centering
    \includegraphics[]{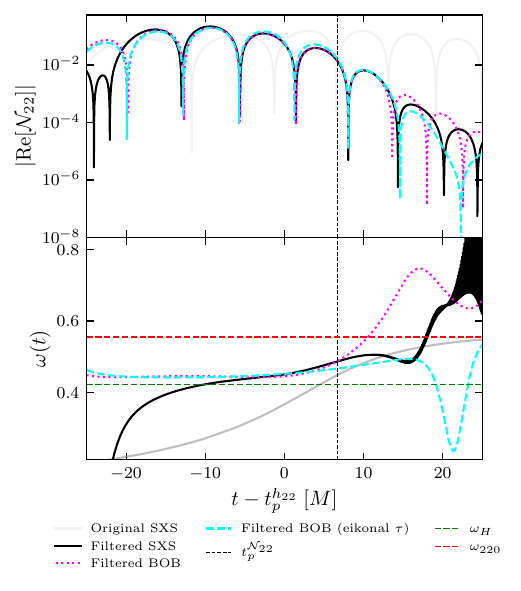}
    \caption{Top: Filtered NR News (SXS:BBH:0305 \cite{SXS:BBH:0305,SXSCatalogData_3.0.0,Boyle2025SXS,sxs_cat3}) waveform (solid black) compared to a filtered BOB waveform with the exact same QNMs removed (dotted red) and an ``eikonal filtered'' BOB waveform (dashed cyan). In light grey we plot the original NR waveform. Bottom: Instantaneous waveform frequency for the filtered waveforms. We also plot the qnm frequency (dashed green) and $w_H = 2\Omega_H$, where $\Omega_H$ is the horizon frequency (dashed red).}
    \label{fig:dw_0305}
\end{figure}

\begin{figure}[h]
    \centering
    \includegraphics[]{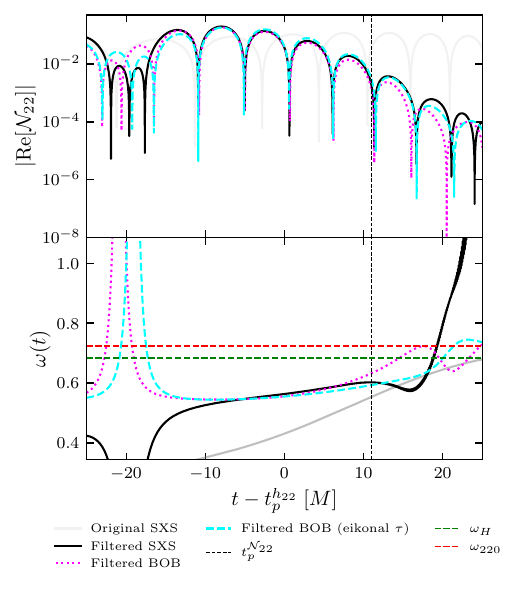}
    \caption{Top: Filtered NR News (SXS:BBH:4123 \cite{Boyle2025SXS,SXSCatalogData_3.0.0,sxs_cat3,SXS:BBH:4123}) waveform (solid black) compared to a filtered BOB waveform with the exact same QNMs removed (dotted red) and an ``eikonal filtered'' BOB waveform (dashed cyan). In light grey we plot the original NR waveform. Bottom: Instantaneous waveform frequency for the filtered waveforms. We also plot the qnm frequency (dashed green) and $w_H = 2\Omega_H$, where $\Omega_H$ is the horizon frequency (dashed red).}
    \label{fig:dw_4123}    
\end{figure}

\begin{figure}[h]
    \centering
    \includegraphics[]{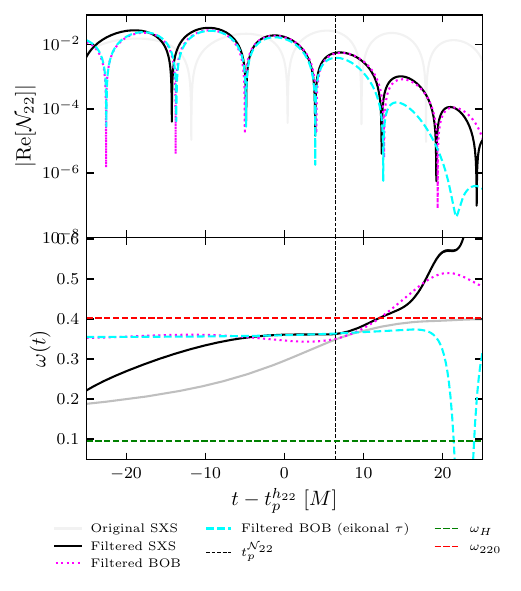}
    \caption{Top: Filtered NR News (SXS:BBH:2477 \cite{Boyle2025SXS,SXSCatalogData_3.0.0,sxs_cat3,Yoo_2022,SXS:BBH:2477}) waveform (solid black) compared to a filtered BOB waveform with the exact same QNMs removed (dotted red) and an ``eikonal filtered'' BOB waveform (dashed cyan). In light grey we plot the original NR waveform. Bottom: Instantaneous waveform frequency for the filtered waveforms. We also plot the qnm frequency (dashed green) and $w_H = 2\Omega_H$, where $\Omega_H$ is the horizon frequency (dashed red).}
    \label{fig:dw_2477}
\end{figure}

In Figs. (\ref{fig:dw_0305}) - (\ref{fig:dw_2477}), we compare filtered BOB waveforms and NR waveforms. While we filter the gravitational wave News, our axis is with respect to the original strain peak for easier comparison to other work. The three SXS waveforms compared here were chosen to span a wide range of initial mass ratios and remnant spins. In all three cases, we find excellent agreement between the filtered NR waveform and the filtered BOB waveforms. We find agreement over a period of $10 - 30 M$, depending on the system parameters and the filtering strategy. Surprisingly, we find that BOB and NR filtered waveforms can agree as far back as $15M$ before the strain peak and up to $20M$ after the strain peak. Furthermore, BOB successfully models this non-QNM component across a wide variety of initial mass ratios and remnant spins. In Figs. (\ref{fig:dw_0305}) - (\ref{fig:dw_2477}), we also plot the waveform frequency for the filtered waveforms. When we use the eikonal limit for the damping time of overtones used in the rational filter, mirroring how BOB incorporates higher overtone information in its amplitude structure, the BOB-filtered frequency shows very little evolution during the period of agreement. Applying the exact same filter to BOB that we apply to the NR waveforms, we find once again that there is very little evolution before the News peak, but after the News peak the BOB filtered frequency shows clear evolution, allowing it to model the filtered NR waveform for longer after the peak.

The authors of \cite{oshita2025probing} interpret the direct wave as prompt emission during plunge, and a model for the direct wave was developed based on the orbital dynamics of a plunging perturber. In the EMRI limit, one expects the perturber to approach the remnant horizon frequency at high spin. 
Based on this, \cite{oshita2025probing} developed a direct wave model that assumes the frequency is near, but not exactly at, the horizon frequency. This approach was used successfully to model the direct wave component of SXS:BBH:0305 in \cite{oshita2025probing} and uncover the direct wave component in GW250114 \cite{abac2025open,abac2026black,abac2025gw250114} in \cite{lu2025gw250114,chung2026measuringblackholesarea}. For SXS:BBH:0305, analyzed in Fig.~(\ref{fig:dw_0305}), the BOB filtered frequencies are near the horizon frequency, consistent with the physical picture in \cite{oshita2025probing}. However, for the high remnant spin system, SXS:BBH:4123, both the NR and BOB filtered waveforms are well below the horizon frequency, and do not approach the horizon frequency until $~20M$ after the strain peak. Because the fundamental QNM frequency $\omega_{220}$ and the horizon frequency are similar for high spins, it is further unclear if the filtered waveform at late times is approaching the horizon or the QNM frequency. For the high mass ratio and low remnant spin system SXS:BBH:2477, analyzed in Fig.(\ref{fig:dw_2477}), the filtered frequency is quite far from the horizon frequency. Through these 3 systems, it is clear that BOB successfully models the direct wave component across initial mass ratios and remnant spins. 

The significant distance of both the NR and BOB filtered frequencies in both low and high remnant spin systems, suggests that the direct wave may be unrelated to the horizon frequency for comparable mass systems. 
To test this systematically, we compute the mean frequency of the ``eikonal filtered" BOB waveform from $5M$ before the peak of $|\mathrm{Re}[\mathcal{N}_{22}]|$ to $5M$ afterwards using a large selection of quasicircular and nonprecessing cases from the SXS catalog \cite{sxs_cat1,sxs_cat2,sxs_cat3}. The dataset spans nearly 600 simulations, up to mass ratio $q$ of 20 and initial spin magnitudes of $|\vec{\chi}_i|\approx 0.99$. We find our results are not sensitive to the exact boundaries of our interval, and the median, maximum and minimum frequency during this period do not show significant deviations from the average quantities, indicating we are correctly identifying the stable frequency in the ``eikonal filtered'' BOB waveform. Additionally, we find the mean frequency of the filtered NR news waveform, removing 7 prograde and 2 retrograde QNMS, over the same time interval. Averaging over the window $[0, +10]$, rather than $[-5, +5]$, raises the NR frequency by 0.02 on average, with a maximum shift of 0.037 across the catalog.

\begin{figure}[h]
    \centering
    \includegraphics[]{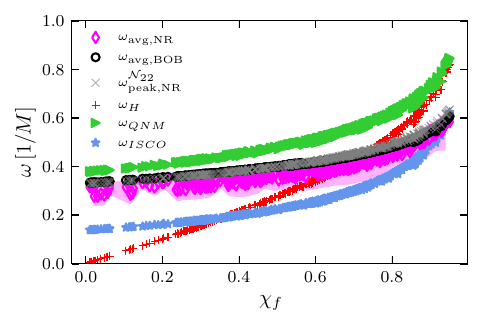}
    \caption{Comparison of the mean ``eikonal filtered'' direct wave frequency for the News (black circle), the mean NR filtered direct wave frequency for the News (pink diamond) with the NR News frequency at the time of the peak News amplitude (grey cross), $\omega_H = 2\Omega_H$, where $\Omega_H$ is the horizon frequency (red plus), real part of the $(s=-2,l=2,m=2,n=0)$ qnm frequency (green triangle) and $\omega_{\text{ISCO}} = 2\Omega_\text{ISCO}$ (blue star). The shaded pink band represents the minimum and maximum values of the NR direct wave frequency in the interval considered.}
    \label{fig:frequencies_sxs}
\end{figure}

Interestingly, Fig.~\ref{fig:frequencies_sxs} shows that the average filtered BOB frequency, $\omega_\text{avg,BOB}$, closely tracks the unfiltered NR News frequency evaluated at the peak of the NR unfiltered News amplitude, $\omega^{\mathcal{N}_{22}}_{\text{peak, NR}}$. The average filtered NR frequency, $\omega_{\text{avg,NR}}$, lies slightly below this value. Moreover, both direct wave frequencies, $\omega_{\text{avg,BOB}}$ and $\omega_{\text{avg,NR}}$, exhibit a different spin dependence than the horizon frequency, $\omega_H$. While they both trend upward with spin, the horizon frequency is a much stronger function of $\chi_f$. Additionally, we observe an incidental crossing around $\chi_f = 0.7$ between the horizon frequency and the direct wave frequency. SXS:BBH:0305, studied in \cite{oshita2025probing} and GW250114, studied in \cite{lu2025gw250114,chung2026measuringblackholesarea}, have final spins of $\chi_f = 0.692$ and $\chi_f = 0.68$ respectively, which explains why direct wave frequency models based on oscillations near the horizon frequency are able to successfully model these two specific systems. However, outside of $\chi_f\approx0.7$, our results suggest that the direct wave frequency may not be a direct probe of the horizon frequency.

\section{Discussion}
In this work we have shown that the amplitude evolution of the Backwards One Body (BOB) model can be derived analytically from black-hole perturbation theory using the Sasaki--Nakamura formalism with the P\"oschl--Teller potential. We demonstrate that the QNM pole contribution resums, to a very good approximation, to a hyperbolic secant envelope. This derivation is based on the observation that the source term $I(\omega_n,r)$ is largely independent of $n$, an insight supported by recent findings obtained from analyzing NR waveforms \cite{mitman2025probing,giesler2019black,cheung2024extracting,oshita2021ease}. The remarkable accuracy of BOB across the SXS catalog for quasi-circular, non-precessing systems further suggests that the source excitation is largely insensitive to the overtone index \cite{kankani2025bob}. Additionally, we show how BOB naturally reproduces the alternating phase structure observed in successive overtones \cite{giesler2019black,ma2022quasinormal,coleman2025high,deamicis2026}.

To understand why BOB can outperform QNM-based models prior to and around the waveform peak, we applied rational filters to both BOB and NR News waveforms. We analyze three SXS cases, a GW150914-like case, a $q=4$ high remnant spin case, and a $q=15$ low remnant spin case. In all 3 cases, we find BOB accurately models the rational filter residual of the merger radiation for an extended period of time, including before the News peak, explaining BOB's accuracy around the waveform peak. BOB's correct incorporation of the QNM pole contributions, in the eikonal limit, \textit{and} its ability to model the direct wave component of the radiation helps explains BOB's ability to outperform a sum of QNMs, on a per-free-parameter basis, near and before the peak of the waveform \cite{kankani2025bob}.

In \cite{oshita2025probing}, based on EMRI waveforms and the GW150914-like SXS:BBH:0305, the direct wave was shown to exhibit a quasi-stable frequency close to, but not exactly at, the horizon frequency for high spin remnants. \cite{lu2025gw250114, chung2026measuringblackholesarea} used a horizon-frequency-based model to extract the direct wave component from GW250114 and associated the recovered signal with properties of the remnant horizon. Similar to \cite{oshita2025probing}, we find that the direct wave has a quasi-stable frequency, with a slow time-dependent evolution. However, we find that this quasi-stable frequency, in both BOB and NR, follows a different spin dependence than the horizon frequency. This suggests that the direct wave may not be a reliable probe of horizon properties. Critically, around $\chi_f\approx 0.7$, the direct wave and horizon frequency happen to coincide, explaining why systems such as SXS:BBH:0305 and GW250114 exhibit a direct wave frequency near the horizon frequency. We find that spin dependence of the direct wave closely tracks the News frequency at the time of the peak News amplitude.

Although BOB was originally motivated by perturbations of null geodesics near the light ring \cite{mcwilliams2019analytical,ferrari}, the direct wave has been associated with radiation emitted by a plunging timelike perturber. However, it is important to note that BOB models the dynamics of the radiation, not the perturber. The radiation emitted just before the light-ring crossing should follow similar dynamics to radiation escaping at the crossing itself. The ability of BOB to accurately model the near merger radiation and the rational filter residual this prompt emission supports the description that the outgoing rays from the plunging source are well modeled by a null-geodesic description. Our results show that BOB effectively captures the QNM and rational filter residual to the merger-ringdown radiation, while requiring only 4 parameters, $M_f,\chi_f,A_p$ and $\Omega_0$. While our analytical results are based on the simpler P\"oschl--Teller potential, the success of BOB for quasi-circular and non-precessing systems across the SXS catalog \cite{kankani2025bob} suggests that the physical intuition obtained from this simpler model extends to the Kerr case. These results, in combination with recent analytical advances in isolating the non-QNM pole contribution \cite{kuntz2025green,su2026decomposition,arnaudo2025beyond}, suggest that connecting BOB analytically to both the direct and QNM pole contributions is a promising direction for future work.

\begin{acknowledgments}
AK thanks Sizheng Ma for helpful discussions and pointing out work on direct waves and rational filters. AK and STM were supported in part by NASA grants 22-LPS22-0022 and 24-2024EPSCoR-0010. The authors acknowledge the computational resources provided by the WVU Research Computing Thorny Flat HPC cluster, which is funded in part by NSF OAC-1726534.
\end{acknowledgments}

\appendix

\section{SN formalism}
Here we provide an overview of the SN formalism, following \cite{berti2006quasinormal}. We take $G=c=2M=1$.
Starting with Eq.~(\ref{eq:sn_inhom_eq}) and assuming the observer is located far from the black hole, we can approximate $\hat{X}^{(s)}(\omega,r)$ as
\begin{equation}
    \hat{X}^{(s)}(\omega,r) \approx \frac{e^{i\omega r_*}}{2i\omega A_{\text{in}}(\omega)} \int_{-\infty}^{+\infty} I(\omega,r_*')\,\hat{X}_{r_+}^{(s)}(\omega,r_*')\,dr_*'
    \label{eq:sn_greens_approx}
\end{equation}
where $\hat{X}_{r_+}^{(s)}$ is subject to the boundary conditions
\begin{align}
    &\lim_{r \to r_+} {\hat X^{(s)}}_{r_+}\sim e^{-i(\omega-m\Omega) r_*}\\
&\lim_{r \to \infty }{\hat X^{(s)}}_{r_+}\sim A_{\text{in}}(\omega)e^{-i\omega r_*}+A_{\text{out}}(\omega)e^{i\omega
r_*}
\end{align}
The subsequent contour integration has to be taken with care to make sure only the QNM contributions are included. The QNM contribution can then be written as 
\begin{equation}
X^{(s)}(t,r)
= -{\operatorname{Re}}\left[\sum_{n}B_n e^{-i\omega_n(t-r_*)}
\int_{-\infty}^{\infty}\frac{I(\omega,r_*'){\hat X^{(s)}}_{r_+}}{A_{\text{out}}}dr_*'\right]
\end{equation}
where $B_n$ are the quasinormal excitation factors, defined as
\begin{equation}
    B_n = \left.\frac{A_\text{out}(\omega)}{2\omega}\bigg(\frac{dA_{\text{in}}}{d\omega}\bigg)^{-1}\right|_{\omega = \omega_n}
\end{equation}
Defining the quasinormal excitation coefficients $C_n$ as 
\begin{equation}
    C_n = B_n \int_{-\infty}^{+\infty} \frac{I(\omega,r_*')\hat{X}_{r_+}^{\text{(s)}}}{A_\text{out}}dr_*'
\end{equation}
we obtain the standard description of the ringdown as a sum of quasinormal modes.
\begin{equation}
X^{(s)}(t,r) =-\operatorname{Re}\left[\sum_{n}C_n e^{-i\omega_n(t-r_*)} \right]
\end{equation}

To relate the master variable $X$ to the gravitational wave strain $h$ as $r\rightarrow \infty$, we follow \cite{Sasaki_strain,berti2006quasinormal} for a non-spinning black hole, $a = 0, s=-2$.
\begin{equation}
    h_+(t) - ih_\times (t) = \frac{8}{r}\sum_{lm} \, _{-2}Y_{lm} (\theta,\phi) \int^\infty_{-\infty} \frac{\hat{X}^{(-2)}}{c_0(\omega)} e^{-i\omega t}d\omega,
\end{equation}
where $c_0=-12i\omega + l(l+1)(l(l+1)-2)$ for $a =0,s=-2$. To go from Eq.~(\ref{eq:dX_dt_propto_sech}) to Eq.~(\ref{eq:news_propto_sech}) we assume the eikonal limit where $c_0 \propto l^2$, although we can see even for $l=2$ the real part is significantly larger. This allows us to move $c_0$ to outside the integral and solve the integral for the PT potential using the same derivation as in the main text. However, for non-spinning systems, we can approximate the leading order correction due to the imaginary part of $c_0$ by treating it in the time domain. Noting that 
\begin{equation}
    \int^\infty_{-\infty} \frac{\hat{f}(\omega)}{c_0(\omega)} e^{-i\omega t}d\omega = \frac{1}{c_0(i\partial_t)}f(t)
\end{equation} 
and
\begin{equation}
    c_0(i\partial_t)[e^{i\alpha \delta t}f(t)] = e^{i\alpha \delta t}c_0(i\partial_t - \alpha \delta)f(t),
\end{equation}
we can use $1/(A+Bx) \approx 1/A(1-Bx/A)$ and Taylor expand $c_0(i\partial_t - \alpha \delta)$ to obtain a leading order correction to the BOB evolution
\begin{widetext}    
\begin{align}
    \mathcal{N}(t) = A_p\text{sech}\bigg(\frac{t-t_p}{\tau}\bigg)\bigg[1 + \frac{2-i\operatorname{Re}(\omega_{220})}{\tau(4 + \operatorname{Re}(\omega_{220})^2)}\tanh\bigg(\frac{t-t_p}{\tau}\bigg)\bigg]e^{i\Phi},
    \label{eq:BOB_correction_app}
\end{align}
\end{widetext}

where $\Phi$ is the original BOB phase evolution and we have replaced the PT QNMs with the full Kerr QNMs. 
Testing across the SXS catalog for quasi-circular and non-precessing systems, we find mixed performance. This is unsurprising considering Eq.~(\ref{eq:BOB_correction_app}) is an approximation of the full non-spinning correction, and we are testing against spinning remnants. In \cite{kankani2025bob}, it was found that BOB showed the worst mismatches for systems with $\chi_\text{eff} \gtrsim 0$ and $q\gg 1$. Interestingly, we find this correction consistently improves BOB's accuracy in this area of the parameter space. Evaluated against 53 cases in the SXS catalog with $q\geq 5$ and $\chi_\text{eff}\geq 0$, we find the mean mismatch decreases from $1.94\times 10^{-4}$ to $7.59\times 10^{-5}$ and the median mismatch decreases from $1.6\times 10^{-4}$ to $4.28\times 10^{-5}$. These results suggest incorporating the full spin effects in the transformation from $X$ to $\mathcal{N}$ may result in corrections to BOB that significantly improve its accuracy across the parameter space.

\bibliography{citations}
\end{document}